\documentclass[useAMS,usenatbib]{mn2e}

\usepackage{graphicx}
\title[Ultraviolet background light and star formation rate]
 {The  ultraviolet extragalactic background light: dust extinction  
   and the evolution of the cosmic star formation rate 
   from $z=0$ to $\sim 0.6$}
\author[J.-M. Deharveng, V. Buat and B. Milliard]{J.-M. Deharveng
\thanks{E-mail:
jean-michel.deharveng@oamp.fr}, V. Buat
and
B. Milliard \\
Laboratoire d'Astrophysique de Marseille, Traverse du Siphon, Les Trois Lucs,
     BP 8, 13376 Marseille Cedex 12, France\\}
\begin{document}

\date{Accepted xxxx. Received xxxx; in original form xxxx}

\pagerange{\pageref{firstpage}--\pageref{lastpage}} \pubyear{2002}

\maketitle

\label{firstpage}

\begin{abstract}
 We show that the accumulated light of galaxies in the ultraviolet can be evaluated from  
 their luminosity density as a function of the evolution of the cosmic star 
 formation rate and dust extinction properties. Constraints on the evolution rate 
 are expected in future. 
 Data available at the moment are consistent with
 an evolution rate at low $z$ steeper than $(1+z)^{3.5}$. A shallower rate 
 remains possible if the luminosity-weighted dust
 extinction at 2000 \AA, as suggested by some data, is lower than $\sim$ 1.2.
\end{abstract}


\section{Introduction}

      The present-day luminosity density of galaxies and evolution effects are folded 
 into the cumulative emission from galaxies. Disentangling these quantities is
 known to be difficult
 because of the need to account for the spectral energy distributions of the galaxies
 at wavelengths shorter than the window of observation. The cumulative emission from
 galaxies may be obtained directly from galaxy number counts or isolated within global 
 measurements of the diffuse background radiation (e.g. \citealt*{ber}).

  In the far-UV wavelength range, the situation is somewhat more simple. 
 The spectral energy distributions are 
 in first approximation dominated by star formation activity,
 avoiding the need to keep track of too many categories of galaxies  
 and opening the possibility of constraints 
 on the history of stellar birth in galaxies. The Lyman
 break enters the window of observation at relatively low redshift, reducing the 
 amount of look-back time involved. The accumulated light of galaxies is the dominant
 source of extragalactic background light which is itself a significant 
 contributor to the total background radiation (e.g. \citealt*{mar,arm}).

  Three factors have recently made a specific examination worth to attempt.
 First, the luminosity density of galaxies is now available at far-UV wavelengths 
 \citep{tre, sul}.
 Second,  the galaxy counts in the UV \citep*{gar} are now deep enough to provide 
 a reliable evaluation of the background radiation due to galaxies, without the difficulties
 of subtracting uncertain  
 components from the observations of the total diffuse background radiation.
 Third, the measurements of the luminosity density of galaxies at H$\alpha$ \citep{gal},
 can be converted into a luminosity density of ionizing photons, then connected by stellar 
 population models
 with the luminosity density above the Lyman break and used as a constraint on 
 the average spectral energy distribution of galaxies in the far-UV.

    In this paper, our motivation is to evaluate the far-UV integrated light of galaxies
 with a limited set of parameters
 and to explore whether useful constraints may 
 be derived on the history of star formation. 
 Despite an explosion in the quantity of faint galaxy data, there are yet 
 significant disagreements in the determinations of the star formation rate (SFR)
 (e.g. \citealt{hop, hog}) 
 that trace the cosmic star formation history.
 Even in relative terms and at low z, the rate of evolution is still a matter of debate, 
 with a parameterization 
 ranging from $(1+z)^{1.5}$ \citep*{cow, wil} to $(1+z)^4$ (e.g. \citealt{lil, mad, mad2}) 
 for measurements based on UV rest frame data, 
 not to speak of more 
 extreme values (e.g. \citealt{hog}) when measurements at other wavelengths are included.

\section[]{Formulation}

    The cosmological radiative transfer equation \citep{pee} for sources with proper 
  specific volume emissivity $\epsilon(\nu,z)$ (in ergs cm$^{-3}$ Hz$^{-1}$ s$^{-1}$)
  gives a mean specific intensity at observed frequency $\nu_0$ and for an observer
  at redshift 0 (e.g. \citealt{bec}),
\begin{equation}
 I_{\nu}= {1\over4 \pi} \int^{\infty}_{0} {dl \over dz} {1 \over (1+z)^3} 
\> \epsilon(\nu,z) \> dz
\end{equation}  
 where $\nu = \nu_0 (1+z)$ and the opacity of the intergalactic medium is 
 neglected for our application in the non-ionizing ultraviolet
 and at relatively low redshift. The relation between the proper length increment 
 and the redshift increment is given by
\begin{equation}
 {dl \over dz} = {c \over H_0} (1+z)^{-1} (\Omega_M (1+z)^3 + \Omega_\Lambda)^{-0.5}
\end{equation}
 where the cosmological parameters and $c$ have their usual meanings.
     In our case the volume emissivity is the ultraviolet emission resulting from 
 the star formation activity in galaxies.  
 Following the notations of \citet{bec}, we assume that it
 can be written as
\begin{equation}
 \epsilon(\nu,z) = \epsilon(\nu_0) (1+z)^3 \psi(z) s(\nu/\nu_0) 
\end{equation}
where $\epsilon(\nu_0)$ is the current local luminosity density of galaxies 
 at frequency
   $\nu_0$, $s(\nu/\nu_0)$ represents the spectral shape normalized
  to 1 at the frequency $\nu_0$ and                   
 $\psi(z)$ accounts for any proper
  evolution in the luminosity density. We take $\nu_0$ as corresponding to 1595 \AA~,
  the pivot wavelength of the far-UV galaxy counts of \citet{gar}. The evolution 
   $\psi(z)$ is currently parameterized as $(1+z)^{\gamma}$. For the purpose of 
  simplification and according to models of star-forming galaxies (continuous star 
  formation), we assume that $s(\nu/\nu_0)$ can be parameterized as 
  $(\nu/\nu_0)^{\alpha}$ between our window of observation and the Lyman break.  

  In these conditions the integrated light from galaxies at 1595 \AA~ is given by 
\begin{equation}
  I_{\nu}(1595) = {c\over4 \pi H_0} \epsilon(1595) \int^{\infty}_{0}  
 {(1+z)^{\alpha+\gamma-1} \over (\Omega_M (1+z)^3 + \Omega_\Lambda)^{0.5}} \> dz
\end{equation}
  In practice the upper bound in equation (4) will be approximately limited to the 
  redshift at which the Lyman break reaches the pivot wavelength of the observations of the 
  integrated light of galaxies \citep{gar}.

\section[]{Evaluation of the UV background radiation due to galaxies}

     We first discuss the two quantities  $\alpha$ and  $\epsilon(1595)$ that,
 in addition to the evolution factor $\gamma$, enter the equation (4) for the calculation
 of the background radiation due to galaxies.

     The normalized spectral shape $s(\nu/\nu_0)$ is written as $(\nu/\nu_0)^{\alpha}
 = (1 + z)^{\alpha}$ under the assumption that the spectral energy distribution 
  results from continuous star formation and dust extinction.
  This assumption is justified
  by the fact that the luminosity density is averaged over a large volume. 
  According to models \citep{lei}, the unreddened s.e.d.
  can be reasonably approximated in the range 1200 \AA~ - 2000 \AA~ by a power-law 
  of slope $\alpha = -0.1$ with current values of IMF
  and metallicity. 
  Such a slope ($-2.1$ for s.e.d. per unit wavelength)
   is observed in star-forming galaxies with 
  low extinction \citep*{cal0}. Models also show that this slope can be extrapolated shortward 
  of 1200 \AA~ but recent observations in the range 900 \AA~ - 1200 \AA~ 
  \citep{lei2, bua} lead us to stop our extrapolation to $\sim$ 1000 \AA~
  because of the many absorption features that depress the flux before the Lyman break
  is effectively reached. In these conditions, the effective slope $\alpha$ is defined
  by the slope between 2000 \AA~ and 1000 \AA. Accounting for the differential dust 
  extinction, the effective slope can be written as,
\[
 \alpha = -0.1 + {{0.4(A_{1000}-A_{2000})} \over {(\log(2000)-\log(1000))}}
\]   
  with $A_{1000}$ and $A_{2000}$ being the amount of extinction at 1000 \AA~ and 2000 \AA~
  respectively. It is finally 
  found to depend on $A_{2000}$ and the ratio of the reddening law at 1000
  \AA~ and 2000 \AA~, $k(1000)/k(2000)$
\begin{equation}  
\alpha  = -0.1 + 1.328 A_{2000} (k(1000)/k(2000) - 1)
\end{equation}
  Following the prescriptions of \citet{cal}
 and their extrapolation shortward of 1200 \AA~ by \citet{lei2}, 
  we have $k(1000)/k(2000)=13.88/8.87=1.565$

   The value of $\epsilon(1595)$ is derived from the 
  luminosity density of galaxies observed at 2000 \AA~ by \citet{sul}, taken as
  $0.84 \times 10^{38}$ ergs s$^{-1}$ A$^{-1}$ Mpc$^{-3}$ for
  H$_0$ = 70 km s$^{-1}$ Mpc$^{-1}$. It is then scaled 
  from $z = 0.15$ to $z = 0$  using the factor $1/(1.15)^\gamma$, and from 2000 \AA~ to 1595 \AA~
  using the slope $\alpha$ determined as a function of $A_{2000}$.

    The integrated light from galaxies, calculated from equation (4),
   is displayed in Fig.~1
  as a function of $A_{2000}$ for three values of the evolution rate. It has been         
  converted into units of photons cm$^{-2}$ s$^{-1}$ A$^{-1}$ ster$^{-1}$ for comparison
  with the range of values 144 -- 195  photons cm$^{-2}$ s$^{-1}$ A$^{-1}$
  ster$^{-1}$ reported by \citet{gar}.  An upper bound of $z = 0.6$ has been used in equation (4)  
  but the integrated light of galaxies would be increased by any
  flux shortward of the Lyman break. 
  The increase has been found to be less than $\sim$ 10\% 
  (even in the case of the evolution rate of $(1+z)^4$) for an upper limit  
  to the Lyman continuum escape fraction of $\sim$ 6\% in the nearby star-forming galaxies
   (e.g. \citealt{hur, hec, deh}).

\section[]{Lyman discontinuity}

     A possible constraint on the parameter $\alpha$ is the need for the luminosity 
  density at 900 \AA~ derived from the observed H$\alpha$ luminosity density
  to be compatible with that observed at 2000 \AA~.
  The Lyman discontinuity predicted by stellar population models 
  is a key element in such a comparison.
   As it is current practice \citep{bru, lei},  
   the Lyman discontinuity is evaluated between 1000 \AA~ and 900 \AA~ 
   in order to avoid the influence 
   of the severe line-blanketing by hydrogen towards the series limit at 912 \AA. 
  The luminosity density at 1000 \AA~ is directly related to that observed at 2000\AA~
  by \citet{sul}, through the intrinsic slope of $-2.1$
  (per wavelength unit) and the extinction adopted $A_{2000}$. 
  Under current conditions, the H$\alpha$ luminosity density observed and corrected for 
  extinction by \citet{gal} (taken as  $1.76 \times 10^{39}$ ergs s$^{-1}$ Mpc$^{-3}$ for
  H$_0$ = 70 km s$^{-1}$ Mpc$^{-1}$) implies a density of $1.3 \times 
  10^{51}$ s$^{-1}$ Mpc$^{-3}$ ionizing photons, and in turn a luminosity density of 
   $0.67 \times 10^{38}$ ergs s$^{-1}$ A$^{-1}$ Mpc$^{-3}$ at 900 \AA~
  (assuming a s.e.d. flat per wavelength unit in the Lyman continuum with no flux 
   below the HeII edge).

    The resulting values of the Lyman discontinuity 
   are listed in Table 1 as a function of $A_{2000}$. Most of these values are larger than 
   the amplitude  of 4 predicted for continuous star formation by current 
   evolutionary synthesis models  \citep{bru, lei}. They 
  would match this prediction for unrealistically low value 
  of the extinction and 
  in a domain of Fig.~1 where 
  the integrated galaxy light would be larger than observed. 

    However, 
   a number of possibilities have been identified that, alone or in 
   combination, would better match the evaluation of the Lyman discontinuity
   with the amplitude predicted by models. 
  \begin{enumerate}
  \item If a fraction $f$ of the ionizing photons escape or, more likely, are trapped 
        by dust before ionization (e.g. \citealt*{ino1,ino2,cha,dop}), 
        the evaluation of the Lyman discontinuity 
        would be multiplied by $f$. It would match the 
        predicted value within the permitted range of background radiation from galaxies
        with a plausible value $f = 0.7$.
  \item The evaluation of the Lyman discontinuity would be decreased if the 
         value adopted from \citet{gal} underestimates the H$\alpha$ luminosity 
         density. Arguments have been given in that sense by \citet{jon} and 
         higher values have been reported \citep{gro, gla}; recently the value of \citet{gal}
         has been revised upward by a factor of 1.6 \citep{per}. 
         The  evaluation of the Lyman discontinuity would be divided by the same factor and 
         would be brought closer to the prediction.      
  \item  The possibility of an  increase of 
        the value predicted by models has also be examined. The amplitude of the stellar
        Lyman discontinuity is rather independent of metallicity and only decreases at very 
        low metallicity \citep{sch}. Recent models including non-LTE effects \citep*{sch2, smi}
        tend to produce smaller discontinuities. In contrast, a relative content 
        in massive stars reduced with respect to the current IMF would 
        increase the discontinuity \citep{lei}. By far, a departure from the assumption 
        of a continuous star formation would be the most radical way for 
        increasing the stellar Lyman discontinuity; we have no possibility to exclude  
        an aging burst contribution in the composite population that would make the 
        stellar Lyman discontinuity larger and consistent with the evaluation. 
  \end{enumerate}

\begin{table}
 \caption{Values of the Lyman discontinuity resulting from observations as a function of 
           the adopted $A_{2000}$ extinction.}
 \label{symbols}
 \begin{tabular}{lccc}
  \hline
     $A_{2000}$  &  $\gamma=1.5$ & $\gamma=2.5$ &  $\gamma=4$ \\
  \hline
     0.  &  4.3  &  3.8 &  3.1 \\
     0.2 &  5.2  &  4.5 &  3.7 \\
     0.4 &  6.3  &  5.5 &  4.4 \\
     0.6 &  7.6  &  6.6 &  5.3 \\
     0.8 &  9.1  &  7.9 &  6.4 \\
     1.0 &  11.  &  9.5 &  7.7 \\
     1.2 &  13.  &  11. &  9.3 \\ 
  \hline
 \end{tabular}

 \medskip
    Values of the Lyman discontinuity are dependent on the evolution rate 
    because the UV luminosity density had to be scaled back to $z \sim 0$ 
   (with the factor  $1/(1.15)^\gamma$) for comparison 
   with the  H$\alpha$ luminosity density.
\end{table}

\begin{figure}
\includegraphics[width=84mm]{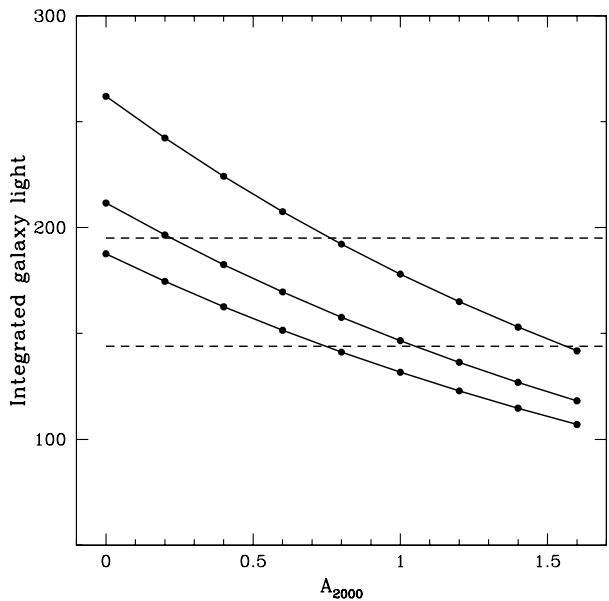}
% \vspace{}
 \caption{Predictions of the integrated light of galaxies at 1595 \AA~ 
 in units of photons cm$^{-2}$ s$^{-1}$ A$^{-1}$ ster$^{-1}$ as a function of the 
 extinction $A_{2000}$ for three values of the evolution rate, $\gamma$ = 1.5, 2.5 and 4 
 from bottom to top (H$_0$ = 70 km s$^{-1}$ Mpc$^{-1}$, $\Omega_M$=0.3, $\Omega_{\Lambda}$
 =0.7).
 The range of values measured by \citet{gar} is indicated as the two
 horizontal dashed lines.}
\end{figure}

\section[]{Discussion and dust extinction}

      If the constraint from the Lyman discontinuity is ignored or assumed to be solved 
   by the effects listed above, the predictions of the integrated light from galaxies
   are compatible with the observations
   for a large range, 
   but relatively low values, of the $A_{2000}$ extinction (Fig.~1). Most of these 
   values are  
   lower than the extinction of 1.3 reported  by  \citet{sul} as affecting
   their luminosity density. Taken at face value, the extinction  $A_{2000} = 1.3$  would 
   imply an evolution rate $\gamma \geq 3.5$;
   a significant adjustment for the Lyman discontinuity
   and the lower bound of the UV integrated light  would be required if $\gamma \sim 3.5 - 4$.
   Such a definite conclusion ($\gamma \geq 3.5$),
   however, is mitigated by a number of arguments.

\begin{enumerate} 

   \item   The $A_{2000} = 1.3$ extinction  compares well with the average 
   UV extinction estimated in starburst galaxies by a number of authors (e.g. \citealt*
   {bua0, meu, cal1, bel}). When it comes to UV selected samples of nearby galaxies,
   lower average extinction has been reported, especially when this 
   extinction is evaluated from the dust emission of the galaxies.
   UV selected samples are expected to contain  
   normal, quiescent star-forming galaxies in addition to starburst galaxies
   as should be  the case for the UV luminosity density discussed here. The average 
   UV extinctions
   reported range from 0.6 to $\sim 1$ \citep*{bua6,bua9,igl}. Such values 
   would be compatible with evolution rate  $\gamma < 3.5$.

   \item The extinction of 1.3 found  by  \citet{sul} for their
   luminosity density is 
   surprisingly close to the average performed on the individual extinctions, whereas a lower 
   value is expected from a luminosity-weighted average.

    \item In addition to the $A_{2000}$ extinction itself and the issue 
   of the  universality of the starburst obscuration curve \citep{bel1}, the ratio
   $k(1000)/k(2000)$ is a factor in the calculation of the integrated light from galaxies
    (cf equation 5). Given the uncertainties on the absorption curve,
   especially below 1200 \AA, a value lower than 
   the adopted 1.565 cannot be ruled out; this would raise the series of curves in Fig.~1
   in the domain of large extinction and would make again the data compatible 
   with a lower evolution rate.

\end{enumerate}
 
\section[]{Conclusion}

   We have tried to put together three quantities deduced from independent measurements, 
 the accumulated far-UV light of galaxies (at $\sim 1595$ \AA),
 the luminosity density of galaxies in the far-UV (at 2000 \AA) and at H$\alpha$.
 As the star formation is likely to be continuous over large volume, we assume that
 the cosmic spectrum, i.e. the luminosity-scaled spectra summed over all galaxies, can be written
 in the far-UV as a simple power-law with an exponent 
 depending only on the dust extinction. In these conditions, the integrated light from galaxies 
 is predicted as a function of the cosmic evolution of the star formation rate
 and average dust extinction. This approach will offer the possibility to pin down 
 the cosmic evolution rate when the UV integrated light from galaxies is better known 
 and the dust extinction better constrained in the UV (as with the GALEX survey).

 Data available at the moment imply an evolution rate 
 at low $z$ steeper than $(1+z)^{3.5}$. A shallower evolution is possible 
 if the properties of 
 the luminosity-weighted dust extinction are less extreme in the UV than those based 
 on starburst galaxies. The latter possibility, 
 with an average extinction  $A_{2000}<1.2$ and/or a reddening law 
 shallower at short wavelengths than predicted by \citet{lei2}, cannot be ruled out and is supported
 by recent trends found in normal galaxies.
 
 The comparison between the luminosity density of galaxies 
 in the far-UV (at 2000 \AA) and at H$\alpha$ is compatible with the stellar Lyman discontinuity 
 predicted by evolutionary synthesis models if a fraction of the ionizing photons are trapped 
 by dust before ionization and/or the H$\alpha$ luminosity density of \citet{gal} is underestimated.

%\section*{Acknowledgments}

%We thank xxxxx for some helpful suggestions,
%xxxxxx for a critical reading of the original version of the
%paper 
%and an anonymous referee for very useful comments that improved
%the presentation of the paper.

\end{document}